\documentclass[twocolumn,showpacs,preprintnumbers,amsmath,amssymb]{revtex4}

\usepackage[dvips]{graphicx}
\usepackage{graphics}
\usepackage{longtable}
\usepackage{stmaryrd}
\usepackage{amssymb}

\begin{document}

\title{First-Principles Method for Open Electronic Systems}
\author{Xiao Zheng}
\author{GuanHua Chen}
\email{ghc@everest.hku.hk} \affiliation{Department of Chemistry,
The University of Hong Kong, Hong Kong, China}

\date{\today}


\begin{abstract}
We prove the existence of the exact density-functional theory
formalism for open electronic systems, and develop subsequently an
exact time-dependent density-functional theory (TDDFT) formulation
for the dynamic response. The TDDFT formulation depends in
principle only on the electron density of the reduced system.
Based on the nonequilibrium Green's function technique,
it is expressed in the form of the equation of motion for
the reduced single-electron density matrix, and this provides thus
an efficient numerical approach to calculate the dynamic
properties of open electronic systems. In the steady-state limit,
the conventional first-principles nonequilibrium Green's function
formulation for the current is
recovered.
\end{abstract}

\maketitle

Density-functional theory (DFT) has been widely used as a research
tool in condensed matter physics, chemistry, materials science,
and nanoscience. The Hohenberg-Kohn theorem~\cite{hk} lays the
foundation of DFT. The Kohn-Sham formalism~\cite{ks} provides the
practical solution to calculate the ground state properties of
electronic systems.
Runge and Gross extended
further DFT to calculate the time-dependent properties and hence
the excited state properties of any electronic
systems~\cite{tddft}. The accuracy of DFT or TDDFT is determined
by the exchange-correlation functional. If the exact
exchange-correlation functional were known, the Kohn-Sham
formalism would have provided the exact ground state properties,
and the Runge-Gross extension, TDDFT, would have yielded the exact
properties of excited states. Despite of their wide range of
applications, DFT and TDDFT have been mostly limited to closed
systems.

Fundamental progress has been made in the field of molecular
electronics recently. DFT-based simulations on quantum transport
through individual molecules attached to electrodes offer guidance
for the design of practical devices~\cite{prllang, prlheurich,
jcpluo}. These simulations focus on the steady-state currents
under the bias voltages. Two types of approaches have been
adopted. One is the Lippmann-Schwinger formalism by Lang and
coworkers~\cite{langprb}. The other is the first-principles
nonequilibrium Green's function technique~\cite{prbguo, jacsywt,
jacsgoddard, transiesta, jcpratner}. In both approaches the
Kohn-Sham Fock operator is taken as the effective single-electron
model Hamiltonian, and the transmission coefficients are
calculated within the noninteracting electron model. It is thus
not clear whether the two approaches are rigorous. Recently
Stefanucci and Almbladh derived an exact expression for
time-dependent current in the framework of TDDFT~\cite{qttddft}.
In the steady-current limit, their expression leads to the
conventional first-principles nonequilibrium Green's function
formalism if the TDDFT exchange-correlation functional is adopted.
However, they did not provide a feasible numerical formulation for
simulating the transient response of molecular electronic devices.
In this communication, we present a rigorous first-principles
formulation to calculate the dynamic properties of open electronic
systems. We prove first a theorem that the electron density
distribution of the reduced system determines all physical
properties or processes of the entire system. The theorem lays
down the foundation of the first-principles method for open
systems. We present then the equation of motion (EOM) for
nonequilibrium Green's functions (NEGF) in the framework of TDDFT.
By introducing a new functional for the interaction between the
reduced system and the environment, we develop further a
reduced-single-electron-density-matrix-based TDDFT formulation.
Finally, we derive an exact expression for the current which leads
to the existing DFT-NEGF formula in the steady-state limit. This
shows that the conventional DFT-NEGF formalism can be exact so
long as the correct exchange-correlation functional is adopted.

Both Hohenberg-Kohn theorem and Runge-Gross extension apply to
isolated systems. Applying Hohenberg-Kohn-Sham's DFT and
Runge-Gross's TDDFT to open systems requires in principle the
knowledge of the electron density distribution of the total system
which consists of the reduced system and the environment. This
presents a major obstacle in simulating the dynamic processes of
open systems. Our objective is to develop an exact DFT formulation
for open systems. In fact, we are interested only in the physical
properties and processes of the reduced system. The environment
provides the boundary conditions and serves as the current source
and energy sink. We thus concentrate on the reduced system.

Any electron density distribution function $\rho(\mathbf{r})$ of a
real physical system is a real analytic function. We may treat
nuclei as point charges, and this would only lead to non-analytic
electron density at isolated points. In practical quantum
mechanical simulations, analytic functions such as Gaussian
functions and plane wave functions are adopted as basis sets,
which results in analytic electron density distribution.
Therefore, we conclude that any electron density functions of real
systems are real analytic on connected physical space. Based on
this, we show below that for a real physical system the electron
density distribution function on a sub-space determines uniquely
its values on the entire physical space. This is nothing but the
analytic continuation of a real analytic function. The proof for
the univariable real analytical functions can be found in
textbooks, for instance, reference~\cite{proof1}. The extension to
the multivariable real analytical functions is straightforward.

{\it Lemma:} The electron density distribution function
$\rho(\mathbf{r})$ is real analytic on a connected physical space
$U$. $W\subseteq U$ is a sub-space. If $\rho(\mathbf{r})$ is known
for all $\mathbf{r}\in W$, $\rho(\mathbf{r})$ can be uniquely
determined on entire $U$.

{\it Proof:} To facilitate our discussion, the following notations
are introduced. Set $\mathbb{Z}^{+} = \{0,1,2,\ldots\}$, and
$\gamma$ is an element of $(\mathbb{Z}^{+})^{3}$, \emph{i.e.},
$\gamma = (\gamma_{1},\gamma_{2},\gamma_{3})\in(\mathbb{Z}^{+})
^{3}$. The displacement vector $\mathbf{r}$ is denoted by the
three-dimensional variable $x = (x_{1},x_{2},x_{3})\in U$. Denote
that $\gamma\,\mbox{!} = \gamma_{1}\,\mbox{!}\:\gamma_{2}\,
\mbox{!}\:\gamma_{3}\,\mbox{!}\,$, $x^{\gamma} =
x_{1}^{\gamma_{1}}\:x_{2}^{\gamma_{2}}\:x_{3}^{\gamma_{3}}$, and $
\frac{\partial^{\gamma}}{\partial x^{\gamma}} =
\frac{{\partial}^{\gamma_{1}}}{\partial x_{1}^{\gamma_{1}}}
\frac{\partial^{\gamma_{2}}}{\partial x_{2}^{\gamma_{2}}}
\frac{\partial^{\gamma_{3}}}{\partial x_{3}^{\gamma_{3}}}$.

Suppose that another density distribution function $\rho'(x)$ is
real analytic on $U$ and equal to $\rho(x)$ for all $x\in W$. We
have $\frac{\partial^{\gamma}\rho(x)}{\partial x^{\gamma}} =
\frac{\partial^{\gamma}\rho'(x)}{\partial x^{\gamma}}$ for all
$x\in W$ and $\gamma\in(\mathbb{Z}^{+})^{3}$. Taking a point
$x_{0}$ at or infinitely close to the boundary of $W$, we may
expand $\rho(x)$ and $\rho(x')$ around $x_{0}$, \emph{i.e.},
$\rho(x)=\sum_{\gamma\in(\mathbb{Z}^{+})^{3}}\frac{1}{\gamma !}
\left.\frac{\partial^{\gamma}\rho(x)}{\partial
x^{\gamma}}\right\vert_{x_{0}} (x-x_{0})^{\gamma}$ and
$\rho'(x)=\sum_{\gamma\in (\mathbb{Z}^{+})^{3}}\frac{1}{\gamma !}
\left.\frac{\partial^{\gamma}\rho'(x)}{\partial
x^{\gamma}}\right\vert_{x_{0}} (x-x_{0})^{\gamma}$. Assuming that
the convergence radii for the Taylor expansions of $\rho(x)$ and
$\rho'(x)$ at $x_{0}$ are both larger than a positive finite real
number $b$, we have thus $\rho(x)=\rho'(x)$ for all $x\in
D_{b}(x_{0})=\left\{x:\left\vert x-x_{0}\right\vert <b \right\}$
since $\left.\frac{\partial^{\gamma}\rho(x)}{\partial
x^{\gamma}}\right\vert_{x_{0}} =
\left.\frac{\partial^{\gamma}\rho'(x)}{\partial x^{\gamma}}
\right\vert_{x_{0}}$. Therefore, the equality $\rho'(x)=\rho(x)$
has been expanded beyond $W$ to include $D_{b}(x_{0})$. Since $U$
is connected the above procedure can be repeated until
$\rho'(x)=\rho(x)$ for all $x\in U$.

We have thus proven that $\rho$ can be uniquely determined on $U$
once it is known on $W$, and are ready to prove the following
theorem.

{\it Theorem:} Electron density function $\rho(\mathbf{r})$ for a
subsystem of a connected real physical system determines uniquely
all electronic properties of the entire system.

{\it Proof:} Assuming the physical space spanned by the subsystem
and the real physical system are $W$ and $U$, respectively. $W$ is
thus a sub-space of $U$, \emph{i.e.}, $W\subseteq U$. According to
the above lemma, $\rho(\mathbf{r})$ on $W$ determines uniquely its
values on $U$, \emph{i.e.}, $\rho(\mathbf{r})$ of the subsystem
determines $\rho(\mathbf{r})$ of the entire system.

Hohenberg-Kohn theorem and Runge-Gross extension state that the
electron density distribution of a system determines uniquely all
its electronic properties. Therefore, we conclude that
$\rho(\mathbf{r})$ for a subsystem determines all the electronic
properties of the real physical system.

The above theorem guarantees the existence of an exact DFT-type method
for open systems. In principle, all we need to know is the
electron density of the reduced system. The electron density
distribution in the environment can be obtained by the analytic
continuation of the electron density function at or near the
boundary. The challenge is to develop a practical first-principles
method.
\begin{figure}
\includegraphics[scale=0.45]{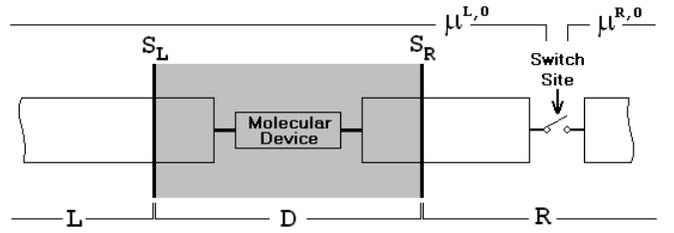}
\caption{\label{scheme} Schematic representation of the
experimental setup for quantum transport through a molecular
device.}
\end{figure}

Fig.~\ref{scheme} depicts one type of open systems, a molecular
device. It consists of the reduced system or device region $D$ and
the environment, the left and right electrodes $L$ and $R$. Taking
this as an example, we develop an exact DFT formalism for the open
systems. To calculate the properties of a molecular
device, we need only the electron density distribution in the
device region. The influence of the electrodes can be determined
by the electron density distribution in the device region.
Within the TDDFT formulation, we proceed to derive the
EOM for the lesser Green's function:
\begin{equation}\label{greenless}
    G^{<}_{nm}(t,t')=i\left\langle
    a^{\dagger}_{m}(t')\,a_{n}(t)\right\rangle,
\end{equation}
where $a_{n}(t)$ and $a^{\dagger}_{m}(t')$ are the Heisenberg
annihilation and creation operators for atomic orbitals $n$ and
$m$ in the reduced system $D$ at time $t$ and $t'$, respectively.
Based on the Keldysh formalism~\cite{keldysh} and the analytic
continuation rules of Langreth~\cite{langreth}, Jauho~\emph{et
al.} developed a NEGF formulation for current
evaluation~\cite{prb94win}. Based on the same procedure adopted in
reference~\cite{prb94win}, we obtain
\begin{eqnarray}\label{eoms4gs}
   i\frac{\partial G^{<}_{nm}(t,t')}{\partial t}
   &=&\sum_{l\in D}h_{nl}(t)G^{<}_{lm}(t,t')\nonumber\\
   &&+\sum_{l\in D}\int_{-\infty}^{\infty}d\tau\Big[\Sigma^{<}_{nl}
   (t,\tau)G^{a}_{lm}(\tau,t')\nonumber\\
   &&+\Sigma^{r}_{nl}(t,\tau)G^{<}_{lm}(\tau,t')\Big],
\end{eqnarray}
where $G^{a}_{lm}(\tau,t')$ is the advanced Green's
function~\cite{prbguo}, $\Sigma^{<}_{nl}(t,\tau)$ and
$\Sigma^{r}_{nl}(t,\tau)$ are the self-energies on $D$ induced by
$L$ and $R$ whose expressions can be found in references such
as~\cite{prbguo} or~\cite{prb94win}, and $h_{nl}(t)$ is the
Kohn-Sham Fock matrix element.
Eq.~(\ref{eoms4gs}) is the exact TDDFT formulation for the open
electronic systems. However, $G^{<}_{nm}(t,t')$ and
$G^{a}_{lm}(\tau,t')$ are the two-time Green's functions. It is
thus extremely time-consuming to solve Eq.~(\ref{eoms4gs})
numerically. Alternative must be sought.

Yokojima \emph{et al.} developed a dynamic mean-field theory for
dissipative interacting many-electron systems~\cite{yokojima}. An
EOM for the reduced single-electron density matrix was derived to
simulate the excitation and nonradiative relaxation of a molecule
embedded in a thermal bath. This is in analogy to our case
although our environment is actually a fermion bath instead of a
boson bath. The major difference is that the number of electrons
in the reduced system is conserved in reference \cite{yokojima}
while in our case it is not. Note that the reduced single-electron
density matrix $\sigma$ is actually the lesser Green's function of
identical time variables,
\begin{equation}\label{1strdm}
    \sigma_{nm}(t)=\left.-iG^{<}_{nm}(t,t')\right\vert_{t'=t}.
\end{equation}
Thus, the EOM for $\sigma$ can be written down readily with the
aid of Eq.~(\ref{eoms4gs}),
\begin{eqnarray}\label{eom4rdm}
    i\dot{\sigma_{nm}} &=& \left.\frac{\partial}{\partial
    t}G^{<}_{nm}(t,t')\right\vert_{t'=t} + \left.\frac{\partial}
    {\partial t'}G^{<}_{nm}(t,t')\right\vert_{t'=t}\nonumber\\
    &=&\sum_{l\in D}\left(h_{nl}\sigma_{lm}-\sigma_{nl}h_{lm}\right)
    +\!\!\!\sum_{\alpha=L,R}\!\!Q_{\alpha,nm}(t),
\end{eqnarray}
where $Q_{\alpha,nm}(t)$ on the right-hand side (RHS) is the
dissipative term due to the lead $\alpha$ ($L$ or $R$) whose
expanded form is
\begin{eqnarray}\label{e4qt}
    Q_{\alpha,nm}(t)&=&i\sum_{l\in D}\int_{-\infty}^{\infty}d\tau\bigg[
    G^{<}_{nl}(t,\tau)\Sigma^{a}_{\alpha,lm}(\tau,t)\nonumber\\
    &&+\,G^{r}_{nl}(t,\tau)\Sigma^{<}_{\alpha,lm}(\tau,t)
    -\Sigma^{<}_{\alpha,nl}(t,\tau)G^{a}_{lm}(\tau,t)\nonumber\\
    &&-\,\Sigma^{r}
    _{\alpha,nl}(t,\tau)G^{<}_{lm}(\tau,t)\bigg].
\end{eqnarray}
And the current through the interfaces $S_{L}$ or $S_{R}$ (see
Fig.~\ref{scheme}) can be expressed as
\begin{eqnarray}\label{jt}
J_{\alpha}(t)&=&
-\frac{d}{dt}\left\langle\sum_{k\in\alpha}c^{\dagger}_{k_{\alpha}}
\!(t)\,c_{k_{\alpha}}\!(t)\right\rangle
    \nonumber\\
&=& \sum_{k\in\alpha}\sum_{l\in D}\left[V_{l,k_{\alpha}}^{\ast}(t)
   G^{<}_{l,k_{\alpha}}(t,t)-G^{<}_{k_{\alpha},l}(t,t)
   V_{l,k_{\alpha}}(t)\right]\nonumber\\
   &=&2\Re\bigg\{\int_{-\infty}^{\infty}d\tau\,\textnormal{tr}
   \Big[G^{<}_{D}(t,\tau)\Sigma^{a}_{\alpha}(\tau,t)\nonumber\\
   &&+\,G^{r}_{D}(t,\tau)\Sigma^{<}_{\alpha}(\tau,t)\Big]\bigg\}
   \nonumber\\
&=& -i\,\mbox{tr}\bigg[Q_{\alpha}(t)\bigg]=-i\sum_{n\in
D}Q_{\alpha,nn}(t),
\end{eqnarray}
where $V_{l,k_{\alpha}}(t)$ is the coupling matrix element between
the atomic orbital $l$ and the single-electron state $k_{\alpha}$
in $L$ or $R$, $G^{<}_{k_{\alpha},l}(t,t')\equiv i\langle
a^{\dagger}_{l}(t')\,c_{k_{\alpha}}\!(t)\rangle$ and
$G^{<}_{l,k_{\alpha}}(t,t')\equiv i\langle
c^{\dagger}_{k_{\alpha}}\!(t')\,a_{l}(t)\rangle$,
$c_{k_{\alpha}}\!(t)$ and $c^{\dagger}_{k_{\alpha}}\!(t')$ are the
annihilation and creation operators for $k_{\alpha}$,
respectively. At first glance Eq.~(\ref{eom4rdm}) is not
self-closed since the $G$s are to be solved. According to the
theorem we proved earlier, all physical quantities are explicit or
implicit functionals of the electron density in $D$,
$\rho_{D}(t)$. $G$s and $\Sigma$s are thus also universal
functionals of $\rho_{D}(t)$. Therefore, we can recast
Eq.~(\ref{eom4rdm}) into a formally closed form,
\begin{equation}\label{e4rdm}
    i\dot{\sigma}=\Big[h[\rho_{D}(t)],\sigma\Big]
    +\sum_{\alpha=L,R}Q_{\alpha}[\rho_{D}(t)].
\end{equation}
Neglecting the second term on the RHS of Eq.~(\ref{e4rdm}) leads
to the conventional TDDFT formulation in terms of reduced
single-electron density matrix~\cite{ldmtddft}. The second term
describes the dissipative processes where electrons enter and
leave the region $D$. Besides the exchange-correlation functional,
the additional universal density functional
$Q_{\alpha}[\rho_{D}(t)]$ is introduced to account for the
dissipative interaction between the reduced system and its
environment. Eq.~(\ref{e4rdm}) is thus the TDDFT formulation in
terms of the reduced single-electron matrix for the open system.
In the frozen DFT approach~\cite{warshel} an additional
exchange-correlation functional term was introduced to account for
the exchange-correlation interaction between the system and the
environment. This additional term is included in $h[\rho_{D}(t)]$
of Eq.~(\ref{e4rdm}). Admittedly, $Q_{\alpha}[\rho_{D}(t)]$ can be
an extremely complex functional. Progressive approximations are
needed for the practical solution of Eq.~(\ref{e4rdm}). Compared
to Eq.~(\ref{eoms4gs}), Eq.~(\ref{e4rdm}) may be much more
convenient to be solved numerically.

To obtain the steady-state solution of Eqs.~(\ref{eom4rdm}) or
(\ref{e4rdm}), we adopt a similar strategy as that of
reference~\cite{qttddft}. As $t,\tau \rightarrow +\infty$,
$\Gamma^{k_{\alpha}}_{nm}(t,\tau)=V_{n,k_
{\alpha}}\!(t)V_{m,k_{\alpha}}^{\ast}\!(\tau)$ becomes
asymptotically time-independent, and $G$s and $\Sigma$s rely
simply on the difference of the two time-variables~\cite{qttddft}.
The expression for the steady-state current is thus as follows,
\begin{eqnarray}\label{sscurrent0}
    J_{L}(\infty) &=& -J_{R}(\infty) = -i\sum_{n\in D}Q_{L,nn}(\infty)
    \nonumber\\
    &=&\int\left[f^{L}(\epsilon)-f^{R}(\epsilon)\right]T(\epsilon)
    \,d\epsilon,\\
    T(\epsilon)&=&2\pi\eta_{L}(\epsilon)\eta_{R}(\epsilon)\nonumber\\
    &&\times\,\mbox{tr}
    \Big[G^{r}_{D}(\epsilon)\Gamma^{R}(\epsilon)G^{a}_{D}
    (\epsilon)\Gamma^{L}(\epsilon)\Big].\label{tofe}
\end{eqnarray}
Here $T(\epsilon)$ is the transmission coefficient,
$f^{\alpha}(\epsilon)$ is the Fermi distribution function, and
$\eta_{\alpha}(\epsilon)=\sum_{k\in\alpha}\delta(\epsilon
-\varepsilon ^{\alpha}_{k})$ is the density of states for the lead
$\alpha$ ($L$ or $R$). Eq.~(\ref{sscurrent0}) is exactly the
Landauer formula~\cite{bookdatta, landauer} in the DFT-NEGF
formalism~\cite{prbguo, jacsywt}. The
only difference is that Eq.~(\ref{sscurrent0}) is derived within
the TDDFT formalism in our case while it is evaluated within the
DFT framework in the case of the DFT-NEGF
formulation~\cite{prbguo, jacsywt}. In other words, {\it the
DFT-NEGF formalism can be exact so long as the correct
exchange-correlation functional is used!} This is not surprising,
and is simply a consequence of that (i) DFT and TDDFT can yield
the exact electron density and (ii) the current is the time
derivative of the total charge.

Just as the exchange-correlation functional, the exact functional
form of $Q_{\alpha}$ on density is rather difficult to derive.
Various approximated expressions have been adopted for the DFT
exchange-correlation functional in the practical implementation.
Similar strategy can be employed for $Q_{\alpha}$. One such scheme
is the wide-band limit (WBL) approximation~\cite{prb94win}, which
consists of a series of approximations imposed on the leads: (i)
their band-widths are assumed to be infinitely large, (ii) their
linewidths $\Lambda^{\alpha}_{k}(t,\tau)$ defined by
$\pi\eta_{\alpha}(\varepsilon^{\alpha}_{k})
\Gamma^{k_{\alpha}}(t,\tau)$ are regarded as energy independent,
\emph{i.e.}, $\Lambda^{\alpha}_{k}(t,\tau) \approx
\Lambda^{\alpha}(t,\tau) \approx \Lambda^\alpha$, and (iii) the
energy shifts are taken as level independent, \emph{i.e.},
$\delta\varepsilon^{\alpha}_{k}(t) \approx
\delta\varepsilon^{\alpha}(t) \approx \delta\varepsilon^\alpha$
for $L$ or $R$.
The physical essence of the transport problem is captured under
these reasonable hypotheses~\cite{prb94win}. In the practical
implementation, the effects of the specific electronic structures
of the leads can be captured by enlarging the device region to
include enough portions of the electrodes.


Following the WBL approximation in reference~\cite{prb94win}, we
obtain that
\begin{equation}\label{efinal}
    Q_\alpha=
    \left[P^{\alpha}(t)-\left[P^{\alpha}(t)\right]^{\dagger}\right]
    -i\left\{\Lambda^{\alpha},\sigma\right\},
\end{equation}
where the curly bracket on the RHS denotes the anticommutator, and
by taking $t=0$ as the switch-on instant $P^{\alpha}(t)$ can be
expressed as
\begin{eqnarray}\label{palpha}
    P^{\alpha}(t) &=& \frac{2}{\pi}\,U^{(-)}
    (t)\Bigg\{\!\int_{0}^{t}d\tau\,\frac{\mbox{e}^{\,i\left(\mu^
    {\alpha\!,\,0}+\delta\varepsilon^{\alpha}\right)(t-\tau)}}
    {t-\tau}\,U^{(+)}(\tau)\nonumber\\
    &&+\!\int_{-\infty}^{0}\!\!d\tau\,\frac{i\,\mbox{e}^{\,i\left[
    \delta\varepsilon^{\alpha}t+\mu^{\alpha\!,\,0}(t-\tau)\right]}}
    {t-\tau}\,G^{r,0}_{D}(-\tau)\!\Bigg\}\Lambda^{\alpha}\nonumber\\
    &&+\,2i\,\Lambda^{\alpha},
\end{eqnarray}
where $\mu^{\alpha\!,\,0}$ is the chemical potential of the lead
$\alpha$ ($L$ or $R$) in its initial ground state,
$G^{r,\,0}_{D}(-\tau)$ is the retarded Green's function of $D$
before the switch-on instant, and $U$s are defined as
\begin{eqnarray}\label{u+u-}
    U^{(\pm)}(t)&=&\exp\bigg\{\pm i\int_{0}^{t}\!h(\tau)d\tau
    \pm\!\!\sum_{\alpha=L,R}\Lambda^{\alpha}t\bigg\}.
\end{eqnarray}
Eqs.~(\ref{efinal})$-$(\ref{u+u-}) constitute the WBL formulation
of the TDDFT-NEGF formalism. Although its explicit functional
dependency is not given, $Q_{\alpha}$ depends implicitly on
$\rho_{D}$ via Eqs.~(\ref{efinal})$-$(\ref{u+u-}).

To summarize, we have proven the existence of the exact TDDFT
formalism for the open electronic systems, and have proposed a
TDDFT-NEGF formulation to calculate the quantum transport
properties of molecular devices. Since TDDFT results in formally
exact density distribution, the TDDFT-NEGF formulation is in
principle an exact theory to evaluate the transient and
steady-state currents. In particular, the TDDFT-NEGF expression
for the steady-state current has the exact same form as that of
the conventional DFT-NEGF formalism~\cite{prbguo, jacsywt,
jacsgoddard, transiesta, jcpratner}, and this provides rigorous
theoretical foundation for the existing DFT-based
methodologies~\cite{langprb, prbguo, jacsywt, jacsgoddard,
transiesta, jcpratner} calculating the steady currents through
molecular devices.

In addition to the conventional exchange-correlation functional, a
new density functional is introduced to account for the dissipative
interaction between the reduced system and the environment. In the
WBL approximation, the new functional can be expressed in a
relatively simple form which depends implicitly on the electron
density of the reduced system. Since the basic variable in our
formulation is the reduce single-electron density matrix, the
linear-scaling techniques such as that of
reference~\cite{ldmtddft} can be adopted to further speed up the
computation.

Authors would thank Hong Guo, Jiang-Hua Lu, Jian Wang, Arieh Warshel
and Weitao Yang for stimulating discussions. Support from the Hong Kong
Research Grant Council (HKU 7010/03P) and the Committee for
Research and Conference Grants (CRCG) of The University of Hong
Kong is gratefully acknowledged.

\end{document}